# A multilingual/multicultural semantic-based approach to improve Data Sharing in an SDI for Nature Conservation[*]

Monica De Martino[1], Riccardo Albertoni[1]

[1]National Research Council
Institute for Applied Mathematics and Information Technology, Genoa, Italy
{Demartino, Albertoni}@ge.imati.cnr.it

## Abstract

The paper proposes an approach to transcend multicultural and multilingual barriers in the use and reuse of geographical data at the European level. The approach aims at sharing scientific terms in the field of nature conservation with the goal of assisting different user communities with metadata compilation and information discovery. A multi-thesauri solution is proposed, based on a Common Thesaurus Framework for Nature Conservation, where different well-known Knowledge Organization Systems are assembled and shared. It has been designed according to semantic web and W3C recommendations employing SKOS standard models and Linked Data to publish the thesauri as a whole in machine-understandable format. The outcome is a powerful framework satisfying the requirements of modularity and openness for further thesaurus extension and updating, interlinking among thesauri, and exploitability from other systems. The paper supports the employment of Linked Data to deal with terminologies in complex domains such as nature conservation and it proposes a hands-on recipe to publish thesauri in the framework.

---

## 1. INTRODUCTION

In the past, the development of an SDI focused mainly on technical interoperability, while problems arising from multilingual and multicultural user groups did not seem very pressing. Nevertheless, in the times of GSDI (GSDI Technical Working Group and contributors, 2004) and INSPIRE (European Commission, 2004) (Bernard et al, 2004) these issues are evidently of particular importance as an SDI now squares up to an international perspective because of users from different countries sharing the same data sources. The INSPIRE directive that entered in force in May 2007 has been adopted by the Commission, with the aim of establishing an infrastructure for spatial information in Europe. Efficient implementation and monitoring require interoperable spatial information across national borders and streamlined access and use of this information by the concerned stakeholders. Thus the development of an SDI must address the multilingual and multicultural issues which increase the complexity of data interoperability. Solutions have to be defined in order to achieve two main objectives: (i) to share geographic information from different sources across Europe despite the cultural difference not only among countries or ethnic groups, but also among communities that have different knowledge domains, (ii) to discover geographic information available at cross-border level despite the multilingual barriers.

These issues require access to geographic data in a standardized way with a common nomenclature. Knowledge Organization Systems (KOS), such as classifications, thesauri, gazetteers, and ontologies, are proposed for sharing standard and scientific terms that are understandable by different user communities operating in the geographic field (Peters et al, 1997). In particular, a shared thesaurus is pivotal for bridging the gap among concepts used in metadata publication and the concepts used in information discovery. However, in a multifaceted domain like nature conservation, different communities with different skills are involved in the management of geographic information and many terminologies are already available to cover the different competencies. A unique nomenclature cannot be provided as a central controlled thesaurus and a flexible solution has to be deployed.

The paper proposes a Common Thesaurus Framework for Nature Conservation. The activity has been performed within the EU project NatureSDIplus (Best Practice Network for SDI in Nature Conservation, ECP-2007-GEO-317007) founded within the eContentplus programme. It covers a cluster of four data themes listed in INSPIRE Annex I and Annex III: Protected sites, Bio-geographical regions, Habitats and biotopes, and Species distribution. The proposed solution is a multi-thesauri framework where different available KOS are assembled in order to provide an integrated terminology for the four data themes. General content thesauri (e.g., GEMET, EARTh) and specific domain resources for the data themes (e.g., EUNIS database on Species and Habitat types) are considered.

Linked Data and semantic web technology are employed to ensure the architectural flexibility required to scale the domain complexity. The proposed framework meets the requirements of openness, modularity, interlinking and exploitability. Modularity and openness guarantee that the framework can be extended and updated with further thesauri. Interlinking among thesauri provides a multicultural flavour: the culturally dependent meanings of a concept can be represented in distinct thesauri to be eventually interlinked. Exploitability is ensured by the thesaurus encoded in a standard and flexible format: it enables third parties applications to exploit the framework content.

The paper makes a multifold contribution:

(a) It fosters Linked Data and semantic web technologies as a fundamental building block to deal with terminologies in complex domains such as nature conservation. In particular, it describes the system requirements and the rationale for the adoption of Linked Data underlining the advantage that such a choice grants in a long term perspective:

(b) It provides a hands-on recipe to publish thesauri according to Linked Data best practice. The recipe is derived from our concrete experience in NatureSDIplus project, which has pioneered the exploitation of Linked Data to share terminological resources in nature conservation;

(c) It describes the Common Thesaurus Framework for Nature Conservation as an available public resource where anyone can get first-hand experience of the linked data advantages.

The paper is organized as following: Section 2 provides an overview of the NatureSDIplus project where the framework has been developed; Section 3 focuses on the principles followed for the framework definition, the design requirements, and the rationale behind the adoption of Linked Data and semantic web technologies. Section 4 illustrates the related work. Section 5 deals with the framework implementation by providing a hands-on recipe to publish thesauri according to Linked Data. Section 6 shows the framework outcome for nature conservation. Section 7 provides a preliminary evaluation of the proposed framework. Section 8 shows the conclusions and future works.

## 2. PROJECT OUTLINE

NatureSDIplus[1] is a Best Practice Network for SDI in Nature Conservation, an EU founded project within the eContent*plus* programme to establish an SDI for nature conservation. The project supports the implementation of the INSPIRE Directive with specific reference to a cluster of four data themes on nature conservation. The considered data themes are: Protected sites (INSPIRE Directive - Annex I); Bio-geographical regions, Habitats and biotopes and Species distribution (INSPIRE Directive - Annex III).

The project aims at enabling and improving the harmonization of datasets on nature conservation, making them accessible and exploitable. In particular, the project outcomes are the design of a metadata profile based on ISO 19115/119 standards, a common data model compliant with INSPIRE specifications and a geo-portal for data access supported by a set of web services.

Multilingual and multicultural issues are dealt with to assure a wider and more effective exploitation of data beside the background of the operator and their location. These issues are addressed by exploiting the thesaurus framework proposed in this paper. The framework aims to be a powerful tool for metadata compilation: for instance, its terms are exploited as controlled vocabulary for compiling the keyword fields of the metadata profile. It is also exploited at the geoportal level: for instance, its terms are exploited by the "auto-completion" services, which provide hints about terms to be used according to some letters typed, supporting the user in resources annotation and query (re)formulation.

## 3. RATIONALE

This section describes the motivations behind the design of the proposed framework underlying the framework requirements and the rationale for Linked Data and semantic web technology employment.

### 3.1. Why a Thesaurus Framework?

Several KOS have been developed in the recent years especially since the European Union has started to address the management of geographic information on a European scale. General purpose KOS (e.g., GEMET[2], EARTh[3]) as well as KOS that pertain more to nature conservation (e.g., CORINE[4], EUNIS[5],

---

[1] http://www.nature-sdi.eu
[2] http://www.eionet.europa.eu/gemet
[3] http://ekolab.iia.cnr.it/earth.htm
[4] http://www.eea.europa.eu/publications/COR0-landcover
[5] http://eunis.eea.europa.eu/

DMEER[6]) are available which were developed by previous initiatives and projects.

Each of these KOS represents only a partial solution covering global or specific aspects for the data themes. Moreover, there is not a clear bibliographic reference summarising the chronological process which has resulted in the finalization of these initiatives and projects as well as making the relationship among their contents clearly understandable and demonstrating their differences. From an analysis we made within the NatureSDIplus consortium, we can state that currently: (i) different communities, which have a large spectrum of competencies, are involved in the treatment and management of information for nature conservation; (ii) many terminologies have been already developed and adopted in the past for covering part of these competencies; (iii) more than one terminology can be available for a given competency; (iv) terminologies adopted often have a national origin, so they are not uniform in all European countries and often even stakeholders from the same country can adopt different terminologies in everyday practice. From these considerations, we think that a brand new thesaurus would be neither pragmatic nor appropriate for addressing the multicultural and multilingual issues. Firstly, it might result in a huge waste of effort attempting to reinvent the wheel. Secondly, another thesaurus placed beside the others could even worsen the current situation by increasing thesauri redundancy. As a consequence, it is more pragmatic to think of the interoperability among the existing vocabularies. Thus we have focused on setting up the best practices required to exploit and complement the efforts already undertaken by third parties.

A proper approach is to create a thesaurus framework which adds and assembles different well known KOS with the intent of providing an integrated view for the different data themes. It has to be a flexible environment where included thesauri may be extended or new thesauri may be added. Figure 1 illustrates the general framework purpose: it is a frame for sharing some general content KOS (e.g., GEMET, EARTh) and specific domain nomenclatures related to some INSPIRE data themes. Each nomenclature is considered to be a sub-thesaurus and it can just be added or linked if necessary with the other thesauri within the framework. In particular, two kinds of inter-thesaurus relationship may be considered: relations between a specific domain thesaurus and a general purpose thesaurus, and relations between two thesauri for different domains. Moreover a thesaurus may be linked with another which is not contained in the framework.

---

[6]ttp://www.eea.europa.eu/data-and-maps/data/digital-map-of-european-ecological-regions

**Figure 1: Schema of the Common Thesaurus Framework for Nature Conservation. BT, RT and NT stand for Broader, Related and Narrower Term**

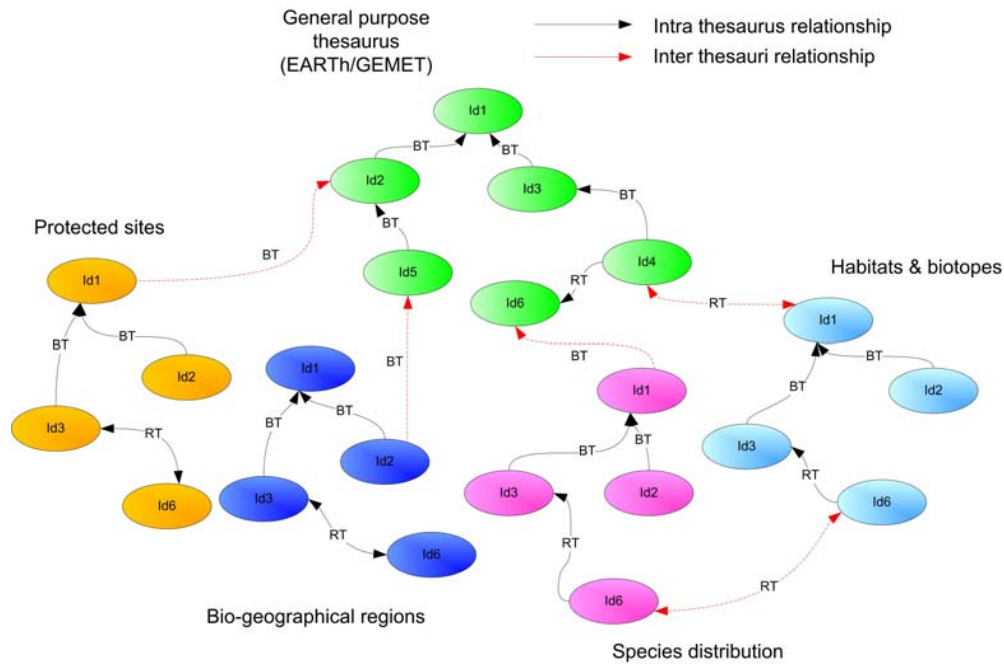

## 3.2. Framework Requirements

Taking the previous considerations into account, we have identified the following key requirements:

- *Modularity*. Each taxonomy/classification/thesaurus should be intended as a module plugged into the set of KOS included in the framework. In particular, modularity should be preserved in order to include updates on existing terminologies;
- *Openness*. Each taxonomy/classification/thesaurus should be easily extendable in order to add (as separated modules) new concepts and terms while keeping the original terminology separated;
- *Exploitability*. KOS should be encoded in a standard and flexible format in order to encourage their adoption and enrichment by third party systems;
- *Interlinking*. Terms and concepts in existing KOS should be interlinked in order to harmonize the term usage. In particular, interlinking is important when a term refers to the same concept in more than one thesaurus because it enables the access to the same concept from a multicultural point of view.

### 3.3. Why Linked Data?

Linked Data (Bizer et al, 2007) and Simple Knowledge Organization Systems (SKOS) (SKOS, 2009) have been selected to meet the aforementioned requirements.

KOS resources are often made available according to distinct models. It is important to point out that they share a similar structure, and are used in similar applications. SKOS captures much of this similarity by providing a common model to represent KOS resources in the Resource Description Framework (RDF) (RDF, 2004). RDF is a very flexible way to structure data; it is a standard model for data interchange on the Web.

Linked Data is a term coined by Tim Berners Lee (Berners-Lee, 2006) which refers to a style of publishing and interlinking RDF structured data on the Web. It has been employed in the thesaurus framework to expose, share, and connect SKOS concepts from different sources. Combining SKOS and Linked Data satisfies the requirements of modularity, openness, exploitability, and interlinking.

The resource translation to SKOS homogenizes the representation of concepts coming from different KOS. It allows inclusion of the sub-thesauri in the framework as separated modules. For each concept, the SKOS model keeps track of the KOS from which the concept has been originated (i.e., concepts from different KOS are included within the framework using different URI namespace). That makes concepts distinguishable according to their origin and therefore makes the task of adding new sub-thesauri easier as well as removing those already included.

Adhering to Linked Data ensures framework openness. A dereferenceable URI is associated with each concept made available in the framework. Thus, third parties can extend the exposed sub-thesauri by referring to their concept URI in the SKOS fragments published everywhere in the Data Web.

Publishing SKOS\RDF according to Linked Data makes the framework remotely accessible it enables both humans and third parties' applications to access and query its content by HTTP and SPARQL ensuring higher exploitability.

Interlinking is achievable thanks to both resource encoding into SKOS\RDF and its availability according to Linked Data. The resource linkage empowers the access to a more complete spectrum of information: it paves the way for exploiting additional information provided by the linked resource.

## 4. RELATED WORK

We have adopted Linked Data (Bizer et al, 2009), SKOS and RDF to develop the thesaurus framework according to the identified requirements.

When the project NatureSDIplus started in October 2008, SKOS was already known as an encoding format for thesauri (Van Assem et al, 2006): for example, it was adopted as an encoding model by GEMET and AGROVOC[7]. The SKOS popularity and maturity were ultimately settled in August 2009 when it became a W3C recommendation (SKOS, 2009). On the contrary, the choice of Linked Data was not settled at all. In 2008, Linked Data was becoming increasingly popular within the semantic web community, but it was not a default choice for publishing thesauri in nature conservation fields. By the time we started using Linked Data to publish the thesauri in the project, encouraging signals had risen from third parties initiatives: GEMET was made available as Linked Data in early 2009.

More recently, in the spring of 2010, EUNIS Species dataset was made available as Linked Data, with a remarkable attempt to relate its species to synonymic species in GBIF[8], ITIS[9], NCBI[10], and WoRMS[11]. EUNIS sites where species were located were also made available and explicit interlinking between species and sites were published. The availability of these KOS throughout Linked Data is extremely exciting, since it offers the possibility of showing how different datasets can eventually be related.

Besides demonstrating these results, our research aims at defining a best practice and providing a simple recipe for publishing existing thesauri for complex domains. It makes content accessible that was previously not available as Linked Data and SKOS (e.g., EUNIS Habitats, and EARTh (Plini et al, 2010)). New interlinking among equivalent or related concepts provided by different thesauri are made available. The aim is to harmonize their usage and to facilitate access to the same concept from a multicultural point of view.

## 5. RECIPE TO SET UP A THESAURUS FRAMEWORK

The multi-step process described below has been followed to build the Common Thesaurus Framework for Nature Conservation:

---

[7] http://aims.fao.org/website/AGROVOC-Thesaurus
[8] http://www.gbif.org
[9] http://www.itis.gov
[10] http://www.ncbi.nlm.nih.gov/guide/taxonomy
[11] http://www.marinespecies.org

1. *Resource selection*: it aims at identifying the available KOS to be included in the framework;

2. *Resource translation to SKOS*: it aims at encoding the selected KOS to SKOS models to import them as sub-thesauri in the framework;

3. *Framework publication*: It aims at making the sub-thesauri available on the web according to Linked Data;

4. *KOS interlinking*: whenever a new sub-thesaurus is published as part of the framework, its connections to already included sub-thesauri are assessed and added.

In the following subsections, the aforementioned steps are more detailed. For each step, the NatureSDIplus experience is discussed as a possible hands-on solution.

### 5.1.    Resource Selection

The first step in the definition of the framework is the identification of a list of available terminological resources.

The selection criteria to be adopted are: (i) relevance of the resource content for the addressed data theme, (ii) availability of the resource translation in different languages, (iii) agreement on resource use (licence policy), (IV) resource availability in digital format (e.g., database or XML).

The resource selection requires the involvement of both technological and domain experts (e.g., stakeholder, data provider). Specific strategies to manage the feedback from the experts must be deployed. The inquiry process is split in two phases:

- *Identification of a preliminary list of candidate resources*. The strategy is to request preliminary feedback about available resources from a large group of experts operating in the data themes;

- *Screening phase* to identify the most appropriate resources contained in the preliminary list. It involves a restricted group of experts and takes into account the aforementioned selection criteria.

In the context of the project, the identification of the candidate resources list has been performed by distributing an on-line questionnaire among the NatureSDIplus partners in order to discover known/used/available KOS for the addressed data themes. At a second stage, the results of the questionnaire were screened by a restricted number of partners with well established expertise in nature conservation. The set of KOS has been selected according to the selection criteria. Table 1 provides the list of the resulting resources and Table 2 provides a view of their characteristics. The list includes a multipurpose thesaurus and at least one sub-thesaurus for each of the four addressed data themes. Resources are in more than one language, and they are available in digital form with established copyright.

**Table 1: List of KOS selected for the framework**

| KOS | Sources | Number of terms | Languages | Data Theme |
|---|---|---|---|---|
| EARTh | CNR | 14340 | en, it | General level |
| IUCN classification | IUCN | 8 | en | Protected sites |
| Habitat types | EUNIS DB | 5431 | en | Habitats and biotopes |
| Nature 2000 A I | | | en, es, nl, el, pt, it, fr, da, fi, sv, de | |
| EUNIS Species | EUNIS DB | 183447 | la, hu, et, pt, es, fr, ro, it, mt, fi, sq, is, da, no, sv, lt, lv, cs, sk, sl, hr, pl, nl, lb, en, bg, ca, de, gr, el, ga, mk, nb, ru, sr, tr, uk | Species distribution |
| Main threats to biodiversity by biogeographic region | EEA | 12 | en | Bio-geographical regions |
| Digital Map of European Ecological Regions | DMEER | About 68 | en | Bio-geographical regions |

### 5.2. Resource Translation to SKOS

The resource translation to SKOS must address two issues:

- Extraction of content from KOS in order to set up the sub-thesauri;
- Translation of the extracted content to SKOS.

In the extraction process, we suggest managing the following information: terms to set up the thesaurus preferred labels (i.e., skos:prefLabel), their descriptions (i.e., skos:definition) and hierarchies to set up concepts definitions and broader\narrower relations (i.e., skos:broader\skos:narrower). We propose keeping the target model as simple as possible: for example, to store data from each resource in a separate table of a relational database, considering distinct columns for each kind of information (e.g., a column for the concept identifier, one for the concept preferred label in Italian, one for the concept preferred label in English, one for the alternative label in English, a column for each relation or interlinking available). MySQL relational database is recommended since it might simplify some of the technical procedures required to map a database table into the SKOS model. If resources are available as databases, store procedures to extract and store each KOS content in an appropriate MySQL table can be employed. If resources are available such as XML, specific procedures must be employed to insert the selected part of information in a MySQL table. If the resources are available in a limited number of short documents, MySQL tables can even be manually compiled. The result is a set of MySQL tables. Each table contains the information properly grouped to be mapped into the SKOS model.

The translation to SKOS is obtained by deploying the mapping between the table columns and the SKOS vocabulary. D2RQ (Bizer et al, 2009) mapping language is suitable for this purpose: it is a declarative language describing the relationships between a relational database and one or more RDFS\OWL schemes [RDFS, OWL]. Since the SKOS structure is defined by an RDFS\OWL schema, D2RQ is even appropriate to formalize the mapping between the MySQL tables and the SKOS model.

In NatureSDIplus the resources were available in different data models.

Table 2 points out their characteristics: none of them were available in SKOS and, except for EARTh, they were classifications or taxonomies. Table 3 shows the SKOS features extracted from each resource.

**Table 2: Characteristics of the resources available for nature conservation**

| Resource | Typology | Data model | Encoding |
|---|---|---|---|
| EARTh | Thesaurus | Relational Database | FireBird |
| IUCN classification | Classification | Document | PDF |
| Habitat types | Taxonomy | Relational Database | Access |
| Nature 2000 A I | Taxonomy | | |
| EUNIS Species | Taxonomy | Relational Database | Access |
| Main threats to biodiversity by biogeographic region | Classification | Document | PDF |
| Digital Map of European Ecological Regions | Classification | Document | PDF |

**Table 3: SKOS features extracted from each resource**

| Resource | skos:prefLabel | skos:altLabel | skos:definition | skos:broader | skos:narrower | skos:related | skos:inScheme |
|---|---|---|---|---|---|---|---|
| EARTh | X | X | X | X | X | X | X |
| IUCN classification | X | | X | X | X | | X |
| Habitat types | X | | X | X | X | | X |
| Nature 2000 A I | X | | X | X | X | | X |
| EUNIS Species | X | X | | X | X | | X |
| Main threats to biodiversity by biogeographic region | X | | | X | X | | X |
| Digital Map of European Ecological Regions | X | | | X | X | | X |

### 5.3. Framework Publication

The D2R Server[12] is proposed to publish the sub-thesauri as Linked Data. Starting from the D2RQ mapping defined in the previous step, the D2R server enables RDF and HTML browsers to access the published content. It also allows third party applications to query linked data employing SPARQL query language (SPARQL, 2008). Figure 2 shows an example of resource translation and publication. It refers to EARTh thesaurus. The implementation is characterized by the following steps:

A. Importing of EARTh DB to the MySQL server. D2R works in principle with any relational database, but some managing facilities are provided for MySQL. For this reason, the first action is to import EARTh DB to the MySQL server;

B. Creation and extraction of a view of the EARTh content mainly because of performance reasons and to simplify the mapping with the EARTh data

---

[12] http://www4.wiwiss.fu-berlin.de/bizer/d2r-server

       model, a view of the EARTh content has been built to implement proper storage procedure;

C. Mapping between the extracted view of EARTh and SKOS. It has been performed by defining a configuration file according to D2R language;

D. Server D2R is setup at a given URL.

**Figure 2: Example of EARTh thesaurus transformation to SKOS and publication**

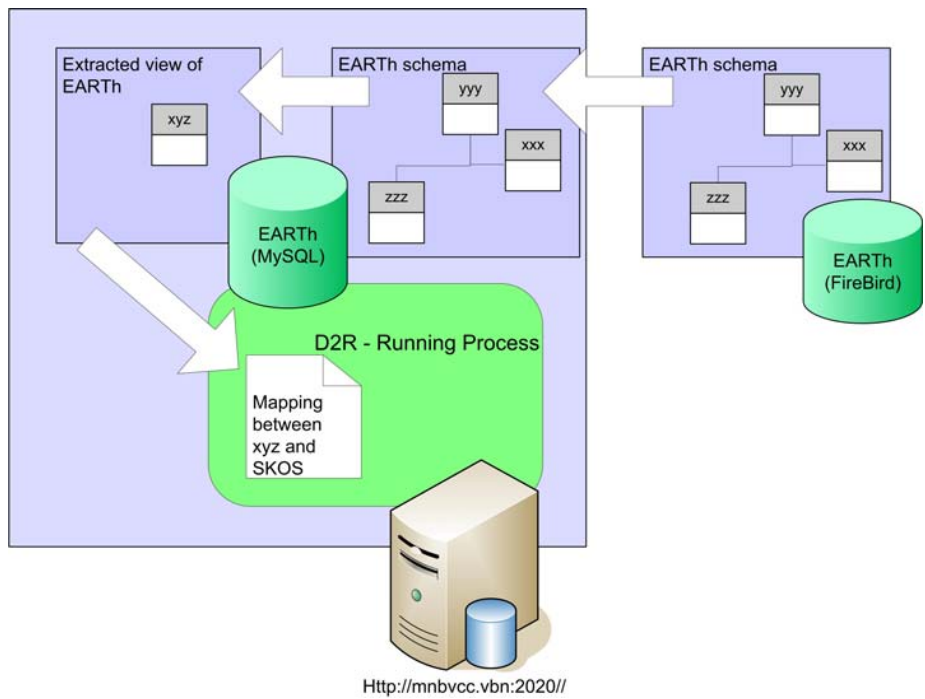

The design choices to represent the SKOS elements with relational database impact on the complexity of mapping. These choices determine what RDF entailments and SKOS constraints can be ensured by exploiting mechanisms which are native in relational database (e.g., primary and foreign keys, integrity constraints). Moreover, an optimal representation depends on the set of design requirements including at least (i) the set of frequent queries; (ii) the frequency of the thesaurus update; (iii) the subset of SKOS integrity constraints and entailments we want to ensure through the relational modelling; (iv) the number and kind of interlinking among sub-thesauri; (v) the number of languages; (vi) the set of SKOS elements to be provided. These design requirements depend on the specific project and so they are not necessarily shared by all sub-thesauri in a Thesauri Framework.

Although an optimal representation of SKOS conceptual model is beyond the scope of this paper, we present two examples of solutions we found in the NatureSDIplus project in order to show how choices in step B impact on step C.

*Example A*. The simplest solution to model the database schema is a scheme where distinct tables for each sub-thesauri are defined and each SKOS element is mapped in a table column. Whenever the SKOS elements are represented in different languages (e.g., for skos:altLabel, skos:note, skos:definition), the available language data is mapped in the column created for that language (e.g., skos_altLabel_It, skos_altLabel_en). Similarly a new column is added to the table for each interlink to subthesauri and external resources.

Given that EARTh is provided with the SKOS elements indicated in the previous Table 3, is available in Italian and English and linked to GEMET and BiogeographicalRegions, then the following table schemas are obtained:

*EARTh (id, prefLabelEn, prefLabelIt, altLabelEn, altLabelIt, descriptionEn, descriptionIT, BT, RT, NT, LinkToGemet, linkToBiogeograficalRegions, inScheme)*

The rationale behind this modeling preference is to keep the D2R mapping file as simple as possible. For each SKOS element there is exactly one column in the table. An excerpt of the mapping file in D2RQ (Bizer et al, 2009) is provided below: the class map (d2rq:ClassMap) represents a class or a group of similar classes of an OWL ontology or RDFS schema, while the property bridge (d2rq:PropertyBridge) relates database table columns to RDF properties. Comment lines starts with '#'.

```
####################
# Table EARTh       #
####################
map:EARTh a d2rq:ClassMap;
        d2rq:dataStorage map:database;
# how to generate the URI of Earth SKOS concepts
        d2rq:uriPattern "EARTh/@@EARTh.ID|urlify@@";
# This skos:Concept
        d2rq:class skos:Concept;
        .
#skos:prefLabel
map:EARTh_ prefLabelEn a d2rq:PropertyBridge;
        d2rq:belongsToClassMap map:EARTh;
        d2rq:property skos:prefLabel;
        d2rq:propertyDefinitionLabel "EARTh TitleEn";
        d2rq:lang "en";
        d2rq:column "EARTh.prefLabelEn ";
        .
#skos:prefLabel
#pretty the same mechanism of prefLabel en changing column and specified language
map:EARTh_ prefLabelIt a d2rq:PropertyBridge;
        d2rq:belongsToClassMap map:EARTh;
```

```
            d2rq:property skos:prefLabel;
            d2rq:lang "it";
            d2rq:propertyDefinitionLabel "EARTh TitleIt";
            d2rq:column "EARTh. prefLabelIt";
            .
#skos:DescriptionEn
…
#skos:DescriptionIT
…
#skos:broader
map:EARTh_BT a d2rq:PropertyBridge;
            d2rq:belongsToClassMap map:EARTh;
            d2rq:property skos:broader;
            d2rq:propertyDefinitionLabel "EARTh BT";
#that defines the URIpatterns for skos concept
d2rq:uriPattern  "http://linkeddata.ge.imati.cnr.it:2020/resource/EARTh/@@EARTh.BT@@";
.
#skos:narrower
…
#skos:related
…
#skos:altLabel en
map:EARTh_SynEn a d2rq:PropertyBridge;
            d2rq:belongsToClassMap map:EARTh;
            d2rq:property  skos:altLabel;
            d2rq:lang "en";
            d2rq:propertyDefinitionLabel " alternative label";
            d2rq:column "EARTh.altLabelEn";
            .
#skos:altLabel it
…
#skos:exactMatch to Gemet
map:EARTh_LinkToGEMEt a d2rq:PropertyBridge;
            d2rq:belongsToClassMap map:EARTh;
            d2rq:property skos:exactMatch;
            d2rq:propertyDefinitionLabel "EARTh euivalence to  GEMETID";
# Starting from a Id contained in Earth we build equivalent Gemet Concept URI
d2rq:uriPattern  "http://www.eionet.europa.eu/gemet/concept?cp=@@EARTh.LinkToGEMEt@@";
.
#skos:inScheme
 map:EARTh_inSchema a d2rq:PropertyBridge;
            d2rq:belongsToClassMap map:EARTh;
            d2rq:property skos:inScheme;
            d2rq:propertyDefinitionLabel "Skos in schemas";
            d2rq:uriPattern
"http://linkeddata.ge.imati.cnr.it:2020/resource/SkosConceptScheme/@@EARTh.inSchema@@"
```

This kind of modelling solution is very simple, but it does not take advantage of the core relational mechanism. It forces modification of the scheme by adding new columns each time a new language and interlinking have to be included. By the way, it provides a swift and convenient solution if a limited number of languages and thesauri are exposed, and these thesauri are not updated very often. Of course other modelling solutions may be more appropriate especially when we want to support the entailments and automatically enforce the integrity constraints derived by SKOS recommendation (SKOS, 2009).

Example B. The database schema illustrated in Figure 3 provides another solution for modelling the database schema in MySQL. The concept scheme and information (e.g. scheme title, description, thesaurus authors, publishers and top concepts) are represented in the tables "SkosConceptScheme", "SkosConceptSchemeInfo" and "hasTopConcept". Each SKOS concept is associated to a skos:ConceptScheme, thus skos:concept(s) are identified by a schema ID (*SkosConcept.skosScheme_ID*) and an intra-scheme ID (*SkosConcept.ID*) in the table "SkosConcept". These identifiers are foreign keys for tables "RDFLabel", "SkosNote" and "SkosSemanticRelation". Moreover: (i) lexical representations of ' skos:concept' such as *skos:prefLabel*, *skos:altLabel* can be stored in the table "RDFLabel" specifying distinct *RDFLabel.lexRelType*; (ii) skos:description and skos:notes can be stored in the table "SkosNote" specifying distinct *skosNote.annotationType*; (iii) semantic relations (e.g. skos:exactMatch, skos:broadMatch, skos:broader, skos:narrower, skos:related) can be stored in the table "SkosSemanticRelation" specifying distinct *SkosSemanticRelation.relType.*

**Figure 3: Database schema in MySQL**

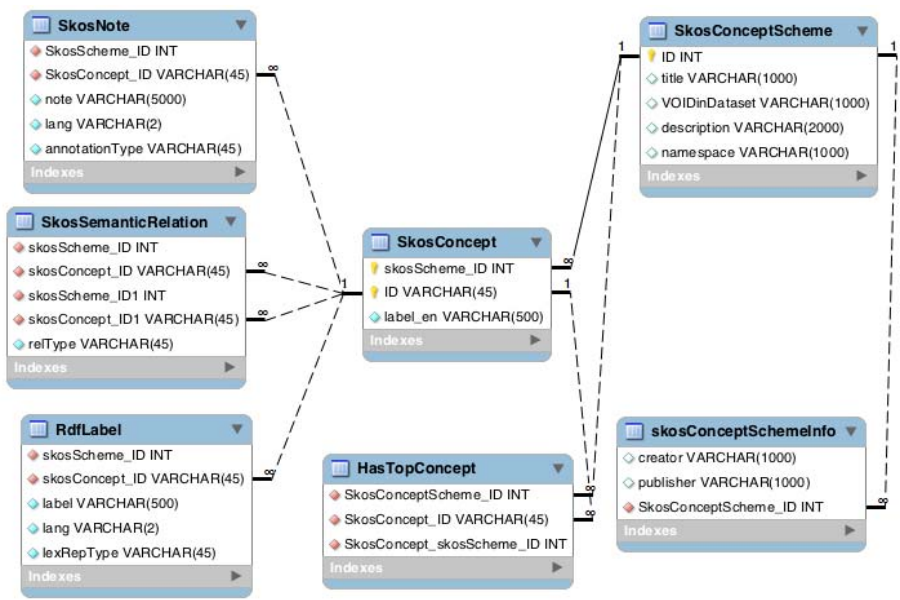

Such a modeling solution is designed to explicity support some of the entailments specified in the SKOS recommendation (SKOS, 2009). In particular, the statement S22 in the SKOS recommendation defines skos:broader as a subproperty of skos:broaderTransitive and S21 defines skos:broaderTransitive as a subproperty of skos:semanticRelation. That means that each skos:broader is also a skos:semanticRelation. Defining the D2R mapping for such a database schema as in the following fragment we can entail that and all the entailments deriving from defined subclass relations between lexical representation and notes.

```
####################
# Table EARTh      #
####################

# Table skosconcept
map:EARTh a d2rq:ClassMap;
        d2rq:dataStorage map:database;
        d2rq:uriPattern "EARTh/@@skosconcept.ID@@";
        d2rq:class skos:Concept;
        #d2rq:classDefinitionLabel " skos concept";
        d2rq:condition "skosconcept.skosScheme_ID=1";
        .
#rdf:label
map:EARTh__label a d2rq:PropertyBridge;
        d2rq:belongsToClassMap map:EARTh;
        d2rq:property rdfs:label;
        d2rq:column "skosconcept.label_en";
        d2rq:lang "en";
        .
#skos:prefLabel (en)
map:EARTh_Titleen a d2rq:PropertyBridge;
        d2rq:belongsToClassMap map:EARTh;
        d2rq:property skos:prefLabel;
        d2rq:column "rdflabel.label";
#  join according to foregn key
        d2rq:join "skosconcept.ID=> rdflabel.skosConcept_ID";
        d2rq:join "skosconcept.skosScheme_ID=> rdflabel.skosScheme_ID";
# consider only pref label in the set of lexical representation
    d2rq:condition  "rdflabel.lexreptype='skosprefLabel'";
    d2rq:condition  "rdflabel.lang='en'";
    d2rq:lang "en";.

#skos:altLabel (en)
map:EARTh_altLabelen a d2rq:PropertyBridge;
        d2rq:belongsToClassMap map:EARTh;
        d2rq:property skos:altLabel;
        d2rq:column "rdflabel.label";
        d2rq:join "skosconcept.ID=> rdflabel.skosConcept_ID";
        d2rq:join "skosconcept.skosScheme_ID=> rdflabel.skosScheme_ID";
        d2rq:condition "rdflabel.lexreptype='skosaltLabel'";
        d2rq:condition "rdflabel.lang='en'";
        d2rq:lang "en";
.
#skos:broader
map:EARTh_broader a d2rq:PropertyBridge;
        d2rq:belongsToClassMap map:EARTh;
        d2rq:property skos:broader;
    d2rq:uriSqlExpression
"CONCAT(skosconceptscheme.namespace,skossemanticrelation.skosConcept_ID1)";
        d2rq:join "skosconcept.ID=> skossemanticrelation.skosConcept_ID";
        d2rq:join "skosconcept.skosScheme_ID=> skossemanticrelation.skosScheme_ID";
d2rq:join "skossemanticrelation.skosScheme_ID1=skosconceptscheme.ID";
d2rq:condition      "skossemanticrelation.relType='skosbroader' OR
# to support entailement w.r.t. S41
skossemanticrelation.relType='skosbroadMatch' ";
```

```
#skos:broaderTransitive
map:EARTh_broader a d2rq:PropertyBridge;
          d2rq:belongsToClassMap map:EARTh;
          d2rq:property skos:broader;
     d2rq:uriSqlExpression
"CONCAT(skosconceptscheme.namespace,skossemanticrelation.skosConcept_ID1)";
          d2rq:join "skosconcept.ID=> skossemanticrelation.skosConcept_ID";
          d2rq:join "skosconcept.skosScheme_ID=> skossemanticrelation.skosScheme_ID";
d2rq:join "skossemanticrelation.skosScheme_ID1=skosconceptscheme.ID";
d2rq:condition       "skossemanticrelation.relType='skosbroader' OR
# to support entailement w.r.t. S41
skossemanticrelation.relType='skosbroadMatch' " OR
# to support entailement w.r.t. S22
skossemanticrelation.relType='skosbroaderTransitive' ";
.
#skos:narrower
... …

#skos:related
… …

# skos:exactMatch
map:EARTh_exactmatch a d2rq:PropertyBridge;
          d2rq:belongsToClassMap map:EARTh;
          d2rq:property skos:exactMatch;
   d2rq:uriSqlExpression
"CONCAT(skosconceptscheme.namespace,skossemanticrelation.skosConcept_ID1)";
          d2rq:join "skosconcept.ID=> skossemanticrelation.skosConcept_ID";
          d2rq:join "skosconcept.skosScheme_ID=> skossemanticrelation.skosScheme_ID";
          d2rq:join "skossemanticrelation.skosScheme_ID1=skosconceptscheme.ID";
          d2rq:condition "skossemanticrelation.relType='skosexactMatch'";
.
# skos:semanticRelation
map:EARTh_semanticrelation a d2rq:PropertyBridge;
          d2rq:belongsToClassMap map:EARTh;
          d2rq:property skos:semanticRelation;
    d2rq:uriSqlExpression
"CONCAT(skosconceptscheme.namespace,skossemanticrelation.skosConcept_ID1)";
          d2rq:join "skosconcept.ID=> skossemanticrelation.skosConcept_ID";
          d2rq:join "skosconcept.skosScheme_ID=> skossemanticrelation.skosScheme_ID";
d2rq:join "skossemanticrelation.skosScheme_ID1=skosconceptscheme.ID";
#there is no condition to support entailement w.r.t. S40, S41, S22, S39, S42
```

Contrary to example A, in this schema, we can store an unlimited number of languages. There is no need to have a distinct table for each thesaurus, and even the interlinking between thesauri is modelled without modifying the database schema. The thesauri that we have exposed are read-only. Users can extend these thesauri, but they are expected to publish such extensions in their own servers. Hence, most of the entailments we have provided are addressed in the storage procedure which imports data from original sources. For example, for every row inducing a skos:broader relation we add a row that materializes its inverse skos:narrower. Incidentally, database triggers can be investigated to support entailments when thesauri are dynamically updated.

### 5.4. KOS Interlinking

Interlinking is one of the hot topics addressed by the Linked Data research community. Different strategies to interlink resources are usually employed. They can be grouped as follows:

*Strategy A: "Exploitation of automatic tools"*. The idea behind automatic tools is to compare concepts belonging to distinct KOS by assessing their similarity, and then to link the concepts whose similarity is higher than a given threshold. Different similarity measures may be employed: they take into account the concept definition and its broader narrower and related concepts. In particular, as pointed out by (Bizer et al., 2009), two research frameworks have been developed by the Linked Data community so far: SILK (Volz et al, 2009) and LinQL (Hassanzadeh et al., 2009). SILK works against local and remote SPARQL endpoints and is designed to be employed in distributed environments. LinQL works over relational databases locally stored. Usually these databases are the back-end of mapping tools such as the D2R server.

*Strategy B: "Exploitation of a priori knowledge"*. Very often KOS are created by common origins or they are built including parts of other pre-existing resources. Knowledge about these interrelations can be crucial to link different KOS, since it suggests a good starting point to find comparable concepts. This strategy is even more effective when two KOS refer to generally accepted naming schemes. For instance, in the domain of nature conservation different agencies have produced data referring to habitat classification defined by the NATURA 2000 network. Whenever each dataset refers to the same identification schemes, the implicit equivalence relationships can easily be made explicit.

*Strategy C: "Exploitation of domain experts"*. The interlinking can be defined manually by domain experts. This activity requires a huge effort. In particular, it is very tricky to reach a consensus especially when a large group of experts are involved. The process can result in a high quality mapping, but only if domain experts are very willing and knowledgeable. In order to attain high quality. It is highly recommended that the experts who have originally defined the thesauri and classifications to be interlinked be involved.

All these strategies have been adopted within our framework.

In the first stage, we have considered SILK and LinQL as automatic tools to interlink the sub-thesauri (strategy A). The rationale behind these tools is very promising and interesting especially from a research point of view. However, they are available only as research prototypes (alpha/beta release). They have not yet reached the technological maturity required for our purposes--to scale up the huge number of concepts that are provided in the framework.

Inspired by LinQL which works at the database level, we have combined the automatic and 'a priori' strategies (strategies A and B) to interlink Habitats and Species of the EUNIS database. Habitats were equipped with descriptions listing the species that they host. Although different taxonomies about species exist, we assumed that the species used in the habitat descriptions were coherent with the species taxonomy since both the resources were provided by EUNIS. Exploiting the 'a priori' knowledge and inspired by the approach followed by LinQL, we have developed a store procedure that adds a link between Species and Habitats whenever some species are mentioned in the habitat descriptions.

The link between EARTh and Bio-geographical regions has been obtained by exploiting the domain experts (strategy C). We have used EARTh developers' knowledge to discover a mapping between EARTh and the resources pertaining to the Bio-geographical regions. This activity was feasible because of the moderate number of concepts contained in the Bio-geographical regions (about 80) and with the help of the expert team who devloped EARTh.

Finally, strategy B has been deployed to interlink EARTh and GEMET, as well as to map EUNIS Species published in our framework to the official Linked Data EUNIS species dataset. In the former, EARTh is a significant extension of GEMET (Plini et al, 2009), and it contains explicit reference to the GEMET ID. Exploiting this reference, it has been possible to link EARTh to GEMET. In the latter, we have exposed the species and their entire taxonomic information (e.g., kingdom, phylum, class, order, family and genus) as SKOS concepts. After our activity, the EUNIS site provided Species as Linked Data. We have interlinked the two datasets: the correspondence between the EUNIS official species and those in the framework was easily identified since we knew how we had exploited the identifiers of the EUNIS species in order to provide an URI for each species in the framework.

## 6. RESULTS

The Common Thesaurus Framework for Nature Conservation includes more than 200,000 concepts covering the four INSPIRE data themes addressed by the project. Most of them come from existing KOS: EUNIS database for species and habitat types, IUCN classification for Protected sites, DMEER and the main threats to biodiversity by biogeographic region classifications for the Bio-geographical regions EARTh is employed as a general purpose thesaurus. It provides about 14,000 concepts, 1023 synonyms and revises and extends GEMET by adding more than 7,000 new concepts. These KOS are in English except for Species which is in Latin; official languages of other EU countries have been partially considered according to the availability of the KOS translations.

Figure 4 illustrates the framework schema showing the KOS modules and their interlinking. In particular, the following interlinks have been generated:

- Habitats to Species and vice versa;
- EARTh to Bio-geographical regions and vice versa;
- EARTh to GEMET;
- Species to official linked data dataset available at the EUNIS/EEA web site.

**Figure 4: The Common Thesaurus Framework for Nature Conservation: the KOS modules and their links**

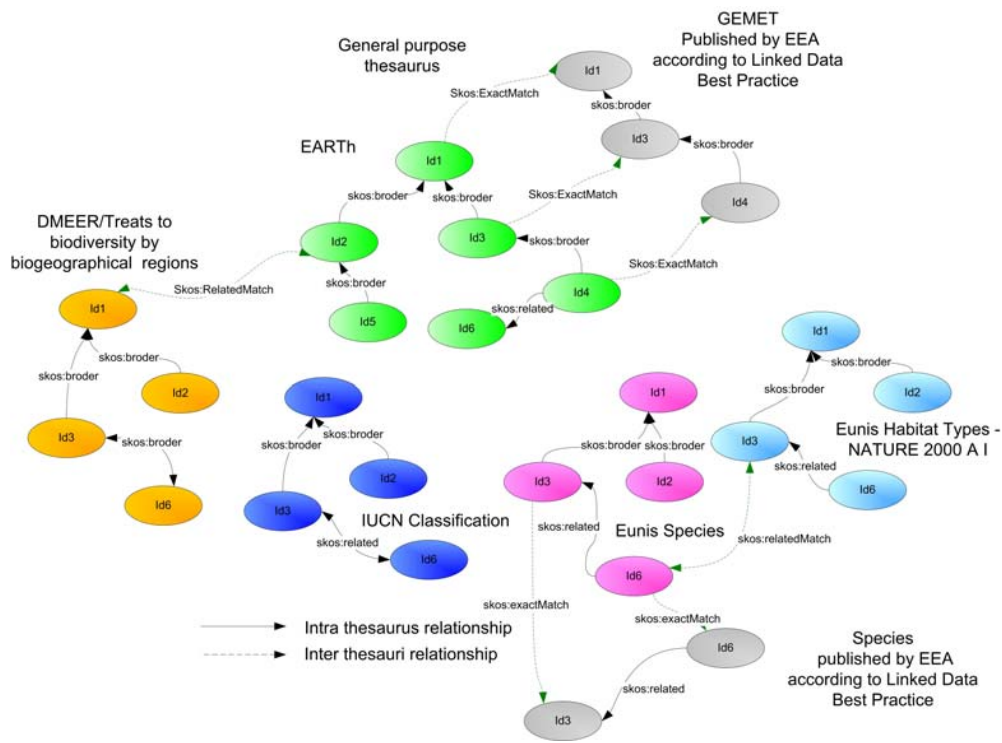

The last two linkages are executed with external resources: GEMET and Species published by EEA according to Linked Data. The resource linkage provides access to a more complete spectrum of information: it paves the way for the exploitation of additional information provided by the linked resources and their future updates. For example, considering EARTh-GEMET linkage, EARTh

226

provides additional concepts and their Italian and English lexical representations to GEMET, whereas GEMET works vice versa in that it extends the multilingual support of EARTh with further translations for the concepts they have in common. Considering the Species linkage, Species within the framework provides the whole taxonomy for Species and their interlinking to Habitats, while the official EUNIS Linked Data provides further useful information such as species synonyms and their vernacular name. Moreover, Species within the framework may exploit the new information which has been published at the official EUNIS Linked Data after the framework implementation (e.g., the interlinking between the species and their pertaining sites).

**Figure 5: Web page of the Common Thesaurus Framework for Nature Conservation**

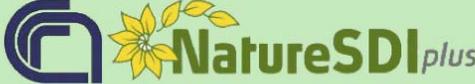

The framework is published at the web site http://linkeddata.ge.imati.cnr.it:2020. Figure 5 illustrates how it looks when accessing it by a web browser. It shows the list of resources available: when an item of the list is selected,, all the pertaining concepts are returned. Moreover,by clicking on one of these concepts, its details become visible: the identifier of the 'broader' (as skos:broader), 'narrower' (as skos:narrow) and 'related' (as skos:related) concepts, concept definition in different languages (as skos:definition) and preferred labels (as skos:prefLabel). Details of broader, narrower and related concepts are obtained by clicking on their respective identifiers.

## 7. PRELIMINARY VALIDATION

This paragraph illustrates a preliminary evaluation of the framework. It aims at demonstrating the following aspects: (i) the suitability of the framework content- the selected resources are representative of the four data themes; (ii) the user vision related to the framework effectiveness in metadata compilation and in information discovery. The effectiveness for metadata compilation and multilingual search cannot be completely evaluated until the NatureSDIplus portal is set up. However, in the meanwhile, we conducted a preliminary evaluation inquiring about the perceived usefulness. Moreover, the evaluation was not used to assess the sub-thesauri contents as well as their interlinking because (i) the framework includes existing sources which have been defined by expert communities and largely adopted in the geographical domain, (ii) the connections among KOS have been created only where they were convenient mainly by exploiting information among concepts stored in the original sources (e.g. the concepts of GEMET which were also in EARTh share the same identifiers, species are stated to characterise the habitats in the description provided by EUNIS).

The evaluation was performed by means of an on-line questionnaire designed with the Surveymonkeys[13] tool and distributed to the project partners. About 22 interviewees answered. Most of them were technological experts or data providers with a good level of expertise in nature conservation and the large majority of them have experience as data providers.

Table 4 summarises the suitability of the framework content. The interviewees agreed with the selected resources with respect to the four data themes. In particular, we can state that: (i) for each theme there is at least one thesaurus which has been highly rated, (ii) looking at the thesaurus rated more than 50%, GEMET supports all themes. In addition to this, the sub-thesauri for the specific themes are assessed as being even more useful than GEMET, (iii) EARTh

---

[13] *www.surveymonkey.com*

seems less appreciated than GEMET, even if it extends GEMET. That is mainly due to the fact the GEMET is more popular than EARTh within the nature conservation community and many interviewees did not know about the inclusion relation between GEMET and EARTh. Table 5 and

Table 6 show the preliminary assessment of the framework usefulness in metadata compilation and data search respectively. In both cases, the percentages of "not useful" are very low (below 16%). There is a tendency to assess sub-thesauri for the data theme as more useful than the general purpose thesauri. Most of the interviewees concluded that the sub-thesauri for the four data themes are useful in data search. That is very encouraging and it goes in the direction of confirming that the framework is useful in metadata compilation and multilingual information search and retrieval. On the other hand, many interviewees responded "do not know" especially for thesauri like GEMET and EARTh. That is an unexpected result, particularly considering that it is common practice to use GEMET and similar thesauri in filling the field keywords of ISO19115 metadata. Interviewees generally ranked the usefulness of thesauri for searching higher than for compiling metadata. That can be interpreted in two distinct ways: (i) they ignore the fact that the thesaurus terms must be inserted in a field of the metadata profile in order to get retrieved in a search; (ii) they think the thesaurus terms have to be extracted by the data and automatically inserted in the metadata. Further investigations about this aspect could be interesting.

Table 4: Percentage of answers to "Please, from your experience indicates which of the following thesaurus/classification can be used for the data themes"

|  | Protected sites | Bio-geographical regions | Habitats and biotopes | Species distribution | None |
|---|---|---|---|---|---|
| EARTh | 30,8% | 38,5% | 61,5% | 53,8% | **23,1**% |
| GEMET | 53,8% | 53,8% | 69,2% | 61,5% | **23,1**% |
| IUCN Classification | **84,6**% | 0,0% | 0,0% | 23,1% | 7,7% |
| Habitat Types & NATURA 2000 A I | 38,5% | 30,8% | **84,6**% | 61,5% | 7,7% |
| EUNIS Species | 15,4% | 23,1% | 46,2% | **84,6**% | 7,7% |
| Main threats to biodiversity by biogeographic region / (DMEER) | 7,7% | **84,6**% | 7,7% | 0,0 | 15,4% |

Table 5: Percentage of answers to "Please indicate if you think it could be useful to exploit the following thesaurus/classification for metadata compilation"

|  | No | Yes | I don't Know |
|---|---|---|---|
| EARTh | 0,0 | 15,4 | **84,6** |
| GEMET | 0,0 | 46,2 | **53,8** |
| IUCN Classification | 7,7 | **61,5** | 30,8 |
| Habitat Types & NATURA 2000 A I | 7,7 | **69,2** | 23,1 |
| EUNIS Species | 7,7 | **53,8** | 38,5 |
| Main threats to biodiversity by biogeographic region/ DMEER | 7,7 | 38,5 | **53,8** |

Table 6: Percentage of answers to "Please indicate if you think it could be useful to exploit the following thesaurus/classification for data search in the geoportal"

|  | No | Yes | I don't Know |
|---|---|---|---|
| EARTh | 15,4 | 30,8 | **53,8** |
| GEMET | 0,0 | **69,2** | 30,8 |
| IUCN classification | 0,0 | **84,6** | 15,4 |
| Habitat types & NATURA 2000 | 0,0 | **92,3** | 7,7 |
| EUNIS Species | 7,7 | **76,9** | 15,4 |
| Main threats to biodiversity by biogeographic region/ DMEER | 0,0 | **53,8** | 46,2 |

## 8. CONCLUSIONS AND FUTURE WORK

The paper proposes a solution for data sharing advancement within an SDI. It refers to an SDI for Nature Conservation in terms of four data themes of INSPIRE Annex I and III (Protected sites, Bio-geographical regions, Habitats and biotopes and Species distribution). It proposes a thesaurus framework which assembles some well-known KOS: it is a powerful tool that satisfies the requirements of openness, modularity, interlinking, and exploitability.

Linked Data combined with SKOS allows framework creation. In particular, this paper encourages the use of Linked Data and semantic web technology to deal with terminologies for complex domains such as nature conservation; it proposes a hands-on recipe for publishing thesauri in a common framework. The adoption of Linked Data enlarges the visibility of the framework both within and outside of the NatureSDIplus consortium identifying it as an example of best practice in the definition and reuse of thesauri.

Moreover, there is currently an advertising activity in progress which is promoting the framework in different communities by defining KOS registries in existing websites, producing VOID descriptions (Alexander et al., 2009) and submitting the KOS to various semantic web search engines such as SINDICE (Oren et al, 2008). This will make the framework available to the worldwide users who are interested in searching for resources with the Data Web.

Future action should explore (i) the extension of the content (by increasing the interlinking among KOS and importing synonyms and terms available in further languages), (ii) the evaluation of how the framework is being used in practice for metadata compilation and data discovery. It can be done as soon as the NatureSDIplus geoportal is available.

It is important to note that the proposed framework is not limited to the nature conservation field, but in principle it can be applied to any geographic domain.

## ACKNOWLEDGEMENT

This activity has been partially supported within the EU project NatureSDIplus (ECP-2007-GEO-317007) of the eContentplus program. Thanks to all partners of the project consortium who have contributed their specific knowledge. We wish to acknowledge Dr Paolo Plini of IIA-ECOLab of CNR in Italy for his cooperation and willingness to furnish the EARTh thesaurus and thanks to ETC/BC for allowing us to access the EUNIS database.